# Modulation of anisotropic magnetoresistance by anomalous Hall signal and its application to real-time domain wall velocity measurement

Ramesh Chandra Bhatt,[1,2] Yo-Yu Cheng,[1,2] Lin-Xiu Ye,[1,2] Ngo Trong Hai,[3] Jong-Ching Wu,[3] and Te-ho Wu,[1,2,a]

[1]Graduate School of Materials Science, National Yunlin University of Science and Technology, Douliu, Yunlin 640 Taiwan ROC
[2]Taiwan SPIN Research Center, National Yunlin University of Science and Technology, Douliu, Yunlin 640, Taiwan ROC
[3]Department of Physics, National Changhua University of Education, Changhua 500, Taiwan ROC

[a]**Author to whom correspondence should be addressed:** wuth@yuntech.edu.tw

## ABSTRACT

We present here a way to modulate the anisotropic magnetoresistance (AMR) by anomalous Hall signal and thus measure the domain wall (DW) motion velocity at near-coercivity. We study the magnetization relaxation at the constant field in the longitudinal ($R_{xx}$) Hall geometry. We observed asymmetric $R_{xx}$ peaks that appear at the DW pinning fields. This unusual magnetoresistance behavior is explained by considering the AMR modulation by the anomalous Hall voltage. In the proposed method, using the magnetization relaxation, the real-time DW velocity measurement is much easier in comparison to the other microscopy methods. Moreover, the additional signal from anomalous Hall voltage makes this technique simpler and sensitive for DW velocity measurements, which can be useful for various spintronic sensing applications.

## INTRODUCTION

The materials with low magnetization such as antiferromagnets and ferrimagnets seem promising for the next generation fast and compact spintronic devices where stray field and limited switching speed, as in the case of ferromagnets, can be avoided. Among various ferrimagnets, the rare earth (RE) - transition metal (TM) alloys have been attracted much attention due to their large perpendicular magnetic anisotropy (PMA). The RE-TM ferrimagnetic alloys possess strong antiferromagnetic exchange between the RE and TM magnetic sublattices, therefore, by changing the content of RE or TM the dominance of these magnetic sublattices can be altered, which at certain composition gives the compensation of magnetization i.e. $M_{net}=0$[1–4]. Moreover, at compensation temperature ($T_{Comp}$), which can be tuned to room temperature, the coercivity ($H_C$) diverges with vanishing $M_{net}$. The non-zero spin polarization and zero magnetic moment at compensation in RE-TM ferrimagnets have the advantages over the antiferromagnets where the magnetic moment, as well as spin polarization, is zero. The easy sensing of magnetization state in these alloys have been investigated for the antiferromagnetic-like ultrafast spintronics applications[5,6].

Recently, the domain wall (DW) motion has been investigated in the thin film of these alloys[7], especially with low current density in TbFeCo strips [8–11]. The DW velocity measurement has been carried out with the help of imaging techniques like the Magneto-optic Kerr effect (MOKE) microscopy and magnetic force microscopy[12] etc. It has been noticed that the DW velocity is measured by the ratio of DW displacement to the current pulse duration[5]. The current pulse may trigger the DW motion but the pulse duration not necessarily be the real-time taken by the DW to travel the distance over the strip, which could give a false large value of DW velocity. An alternate method for DW velocity measurement from anomalous Hall measurements has also been applied, where the DW is created by an additional transverse writing current pulse in addition to the out-of-plane field in Co/Ni films[13]. In another study, *Hayashi et al.* studied the current influence on field-driven DW motion through anisotropic magnetoresistance in permalloy nanowire[14]. There, an in-plane driving field is applied opposite to the current direction and a current pulse is applied in one pickup line to start the DW propagation. The resistance change due to AMR is too low to sense the change in the magnetization state ($\Delta R \sim 0.2\ \Omega$) concerning the overall resistance of nanowire ($\sim 470\ \Omega$) therein, which needs additional arrangements. We apply a rather different approach for measuring the DW velocity, which combines the anisotropic magnetoresistance (AMR) and anomalous Hall effect (AHE) in perpendicular magnetic anisotropic TbFeCo. Therefore, we get an enlarged signal from the modulation of AMR by AHE voltage without any additional arrangements. Here, the DW velocity is defined as the real-time propagation of domain wall in the Hall bar during the magnetization switching from one antiparallel state to another. In the present study, the domain wall propagates at fixed pinning fields whereas in other studies the current pulse with a magnitude of critical value or magnetic field larger than the coercivity is applied that gives higher DW velocity or faster magnetization switching [7].

Here we measure the domain wall velocity in the TbFeCo Hall bar pattern from the magnetization relaxation in the longitudinal Hall geometry. In transverse Hall geometry, the measured resistance is perpendicular to the current direction known as anomalous Hall resistance ($R_{xy}$), whereas, in longitudinal Hall geometry, resistance is measured parallel to the current direction which typically originates due to the anisotropic magneto-resistance (AMR) and here denoted as $R_{xx}$. A constant and perpendicular DW depinning field is applied and then change in longitudinal resistance is measured as a function of time. The origin of asymmetric magneto-resistance in longitudinal Hall geometry is explained and the domain wall motion velocities have been estimated from the magnetization relaxation curves at different magnetic fields. We present here a simple and unique method to obtain the domain wall motion velocity and its distribution over the Hall bar channel.

**MATERIALS AND METHODS**

The TbFeCo amorphous alloy film with structure Hf(4nm)/TbFeCo(10nm)/MgO(2nm)/Ta(2nm) was deposited on thermally oxidized Si-substrates using magnetron sputtering (base pressure 2.2 x $10^{-7}$ Torr). The TbFeCo was co-sputtered from the Tb and $Fe_{80}Co_{20}$ targets so that the composition of the alloys can be changed by changing the Tb target power. The magnetic properties of the deposited film were measured using an Alternating-Gradient Magnetometer (PMC AGM) at room temperature. The Hall bar device for the anomalous Hall measurements was fabricated by electron-beam lithography and Ar-ion milling. The anomalous Hall measurements were carried out using an electromagnet-equipped four-probe station at room temperature.

## RESULTS AND DISCUSSIONS

The schematic of multilayer structures and device connections are shown in Fig. 1(a). The Hall voltage pickup line widths are 5 and 1 µm and the current channel width is 5 µm in the Hall bar device. The in-plane and out-of-plane magnetic hysteresis loops for blanket-film which were measured by an AGM are shown in Fig 1(b). The saturation magnetization and the coercivity of the blanket film are measured to be 94 emu/cc and 770 Oe, respectively. The hysteresis loops show the perpendicular magnetic anisotropy in the multilayer structure. As shown in Fig. 1(c), both up and downward $R_{xx}$ peaks coincide with the $R_{xy}$ coercivity. The lower value of coercivity i.e. 567 Oe for the device is possible due to the patterning process. In general, $R_{xx}$ peaks appear symmetric irrespective of the sweeping field polarity with magneto-resistance ($\Delta R/R$) of $10^{-2}$-$10^{-4}$ order, which is known as anisotropic magnetoresistance (AMR) [15]. However, in the present case, the change in $R_{xx}$ is much larger and the peaks appear asymmetric with respect to the sweeping field polarity. This unusual magnetoresistance will be discussed in the later section of this manuscript. Here we performed the magnetization relaxation for $R_{xx}$ configuration and analyzed the magnetization relaxation dynamics therein.

For performing magnetization relaxation first a perpendicular saturating field of 1 kOe is applied for a while to generate a single domain state. A perpendicular depinning field in the opposite direction is then applied across the Hall bar. The moment when the depinning field reached to desired value the longitudinal resistance ($R_{xx}$) is measured between the two voltage pickup lines by applying a 100 µA dc sensing current along the current channel as shown in Fig. 1(a). All measurements were carried out at room temperature. The $R_{xx}$ vs $H_z$ plot, where the $R_{xx}$ is measured between the two pickup lines, is shown in Fig. 2(a). The magnetization relaxation curves measured in various fields are shown in Fig. 2(b). The valley of the magnetization curve shifts to lower time with an increase in the pinning field. The width of the curve also decreases with an increase in the pinning field, which indicates faster switching. To interpret the domain wall motion in the device each magnetization curve has been analyzed in detail.

The longitudinally measured magnetization relaxation curve for $H_z$= 485 Oe is shown in Fig. 3, where the distance traveled by domain wall and the DW velocity distribution over time is also shown. As we described before, the asymmetric peaks coincide with the coercivity, therefore, we measured $R_{xx}$ as a function of time at a constant driving field to take advantage of large signals from the change in $R_{xx}$. Usually, the MR peaks appear symmetric (i.e. both peaks up), the origin of asymmetric (i.e. up and down) magnetoresistance peaks in the present case needs to be explained to better understand the DW motion.

The magnetoresistance anomaly has been observed before in the DyFeCo thin films where the magnetic moment canting between the sublattices and the magnetization divergence at the edges has been accounted for its origin[16]. The origin of asymmetric magnetoresistance behavior is explained by a schematic of domain wall motion as shown in Figs. 3(f-h). Initially as shown in Fig. 3(f), when the device is saturated in either direction the voltage pickup lines are at the same potential and the resistance change is according to the anisotropic magnetoresistance (AMR). However, when the magnetic field is enough to start domain wall motion, the DW moves from the left end to the right end of the voltage pickup line A and the voltage at this pickup line reverses from the initial value as shown in Fig. 3(g), therefore, there exists a maximum change in the potential between A and B pick up lines.

When DW propagates by the driving field it crosses the pickup line B and the voltage at pickup line B is reversed from the initial value in Fig. 3(h), therefore, again both the pickup line A and B are switched to the same potential and the resistance returns to the initial value. The initiation of DW propagation from the left edge could be due to the inhomogeneity, which allows nucleation at much lower driving fields. One can think about the possibility of multiple domain propagation in the device, however, it has been observed from the MOKE imaging that the TbFeCo sheet film exhibits single domain propagation on increasing the driving field. Moreover, the presence of multiple domains would have shown different shapes in the $R_{xx}$-$H_z$ curve, however, the repetitive measurements at different fields show the same curve shape and the signal strength as well, rule out the multiple domain possibility.

The AMR is extremely small in comparison to the anomalous Hall resistance and is always symmetric i.e. the AMR peaks are always positive. Therefore, a slight difference in the height of the positive and negative peaks appearing in $R_{xx}$ vs field or time plot can be seen which is equal to the ($R_{xy}$ + $\Delta R_{AMR}$) and (-$R_{xy}$ + $\Delta R_{AMR}$), where, $\Delta R_{AMR}$ is the change in resistance due to the AMR. The difference in the height of the peaks for different sensing currents can also be seen in Fig. 4 (a). The $R_{xx}$ curve-peaks for 3 mA current are shifted to lower coercivity and a vertical shift in $R_{xx}$ at zero field is noticed, which is possible due to the Joule heating effect from the large current. Now, for crossing the pickup line B, DW needs to travel the distance between A and B plus the width of B, which is larger than the distance traveled by DW crossing the width A. Therefore, DW takes more time to cross the width B, which is evident from the magnetization relaxation curves.

For analysis of the magnetization relaxation curves, the relaxation time curve can be divided into two parts by a verticle line passing trough the $R_{xx}$ minima as shown in Fig. 3(c). These two parts left-side and right-side are related to the two voltage pickup lines A (width 5 μm) and B (width 1 μm), respectively. The microscopic image and schematic of the Hall bar pattern are shown in the panel of Fig. 3(c). For DW velocity measurement, the resistance was converted to the equivalent distance traveled by the DW. In Fig. 3(a), distance 0 to 5 μm corresponds to the width of pickup line "A" with the origin at the left end of the pickup line "A". In Fig. 3(b), distance 0 to 31 μm corresponds to the sum of the distance between pickup lines and the width "B" (i.e. 30 μm + 1 μm) with the origin at the right end of the pickup line "A". The second half corresponds to the distance 31 μm as the magnetization is relaxed completely to the initial magnetization/resistance value. The $R_{xx}$ around 622 Ω corresponds to the initiation of domain wall propagation at the left end of line "A" whereas the $R_{xx}$ around 618 Ω corresponds to the right end of line "A" where reversal of Hall voltage takes place by the propagation of domain wall. Similarly, in the left dissected part 618 Ω and 622 Ω correspond, respectively, to the right end of line "A" and right end of the line "B". In the left dissected part the time is considered up to the point where the relaxation curve saturates. The initial (at 0 times) or final position of the relaxation curve has been normalized to the corresponding distance. The velocity vs time plot is obtained by taking the first derivative of distance vs time plot. Such distance vs time plots for the two halves of the magnetization relaxation curve is shown in Fig. 3(a) and 3(b), respectively. The first derivative of these plots gives the DW velocity distribution which is also shown in Fig. 3(d) and 3(e). The average DW velocity is estimated by the slope of the linear fit to the distance vs time plot. The estimated average DW velocities are 0.33 μm/s and 1.77 μm/s when crossing the pickup line A and B, respectively.

The DW velocity distribution indicates the maximum velocity peaks of ~ 0.84 µm/s and ~ 3 µm/s for pickup lines A and B, respectively. It is evident from the velocity distribution curves that the full width at half maximum (FWHM) is larger for the domain motion in the left channel.

The distribution of the peak value of domain wall motion velocities with applied perpendicular fields is shown in Fig. 4(b). The DW velocity increases with the field for DW crossing the pickup line B, however, for DW crossing pickup line A the velocity does not influence much by increasing the magnetic field. The DW velocities at line A are much smaller than the velocities when DW crossing line B. We think that this difference could be due to the slow initiation of the DW at pickup line A, where the width of the pickup line is 5 times larger than the width of line B. The difference in pickup line areas allow detouring of DW at line A, which limits the DW velocity. This effect is small at line B. Further, the detouring of DW and inhomogeneity at the left edge could be the possible reason for the unchanged velocity on increasing the field at line A.

**CONCLUSIONS**

In summary, we studied the magnetization switching in the longitudinal i.e. $R_{xx}$ Hall bar geometry in Hf/TbFeCo Hall bar device. We observed that the asymmetric peaks in $R_{xx}$ vs field curve appear at the coercivity of anomalous Hall loop. We measured the domain wall motion velocity at various DW pinning fields through the magnetization relaxation while measuring the $R_{xx}$. The unusual magnetoresistance behavior is explained as a combined effect of the anomalous Hall effect and the anisotropic magnetoresistance. The peak velocity reached a maximum value of around 13 µm/s in the device. The smaller domain wall motion velocity (~ 10 µm/s) in comparison to the reported large values (~ 10 m/s) attributes to the domain wall propagation at small DW pinning fields and real-time duration taken for the magnetization switching in the present case. This study demonstrates an easy way to measure domain wall motion velocity from the magnetization relaxation in comparison to the other complex methods which could be useful in spintronic sensing applications particularly where MOKE imaging is not possible.

This work was supported by the Ministry of Science and Technology (MOST) Taiwan ROC (Grant Nos. MOST 105-2112-M-224-002 and MOST 107-2112-M-224-001-MY2) and the Feng-Tay foundation Taiwan ROC. The author RCB would like to acknowledge the Feng-Tay foundation Taiwan ROC for funding his post-doctoral research.

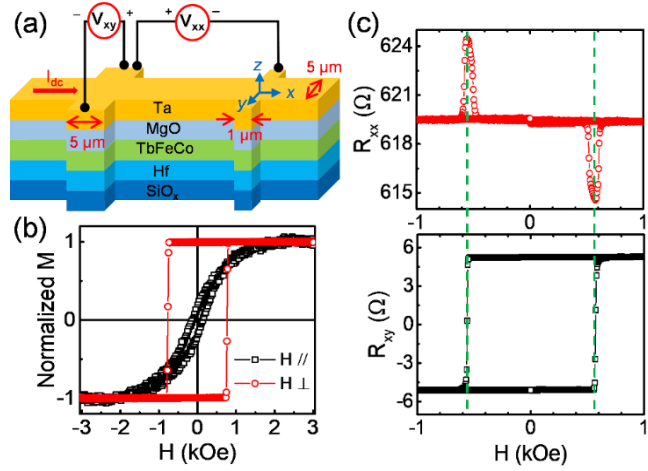

**FIG. 1.** (a) Represents the schematic (not to scale) of thin film multilayer structure and electrical connections of the device. The $V_{xy}$ voltage pickup line widths are 5 and 1 μm and the current channel width is 5 μm. (b) Shows the in-plane and out-of-plane magnetic hysteresis loops measured for the blanket film. Panel (c) represents the longitudinal ($R_{xx}$) and transverse (anomalous Hall resistance $R_{xy}$) resistance as a function of perpendicular field for the Hall bar device. The peaks in $R_{xx}$ curve coincide with $R_{xy}$ coercivity.

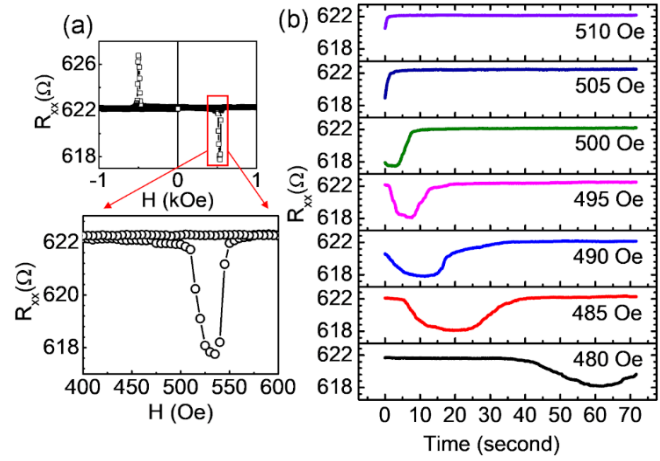

**FIG. 2.** (a) The longitudinal resistance ($R_{xx}$) measurement with field scanning and magnified view of the MR peak for positive field scanning. (b) Magnetization relaxation curves in longitudinal Hall geometry at various perpendicular fields. The magnetization relaxation curve as a function of time mimics the $R_{xx}$ vs field curve indicates the domain wall motion with time.

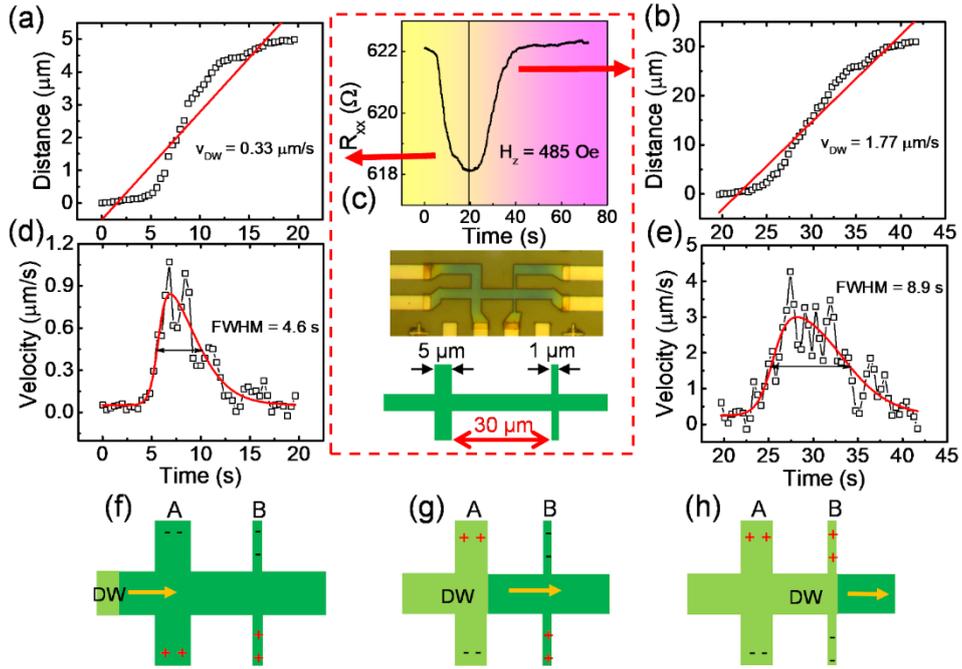

**FIG. 3.** Domain wall motion velocity from the magnetization relaxation in longitudinal Hall geometry at constant perpendicular field. (a) and (b) show the distance traveled by domain as a function of time for left and right-side part of $R_{xx}$-curve shown in panel (c), respectively. The top panel (c) figure shows magnetization relaxation at drive field 485 Oe, where the curve has been dissected vertically into two parts at curve minima. The bottom panel (c) figures show Hall bar pattern and corresponding pickup lines width and distance between them. (d) and (e) show domain wall motion velocity distribution as a function of relaxation time for (a) and (b), respectively. The red solid line represents the linear and nonlinear curve fitting. Representation of domain wall motion in Hall bar device is shown in (f-h). (f) Initiation of DW motion, (g) DW crossing the voltage pick-up line A and (h) DW crossing the second voltage pick-up line B.

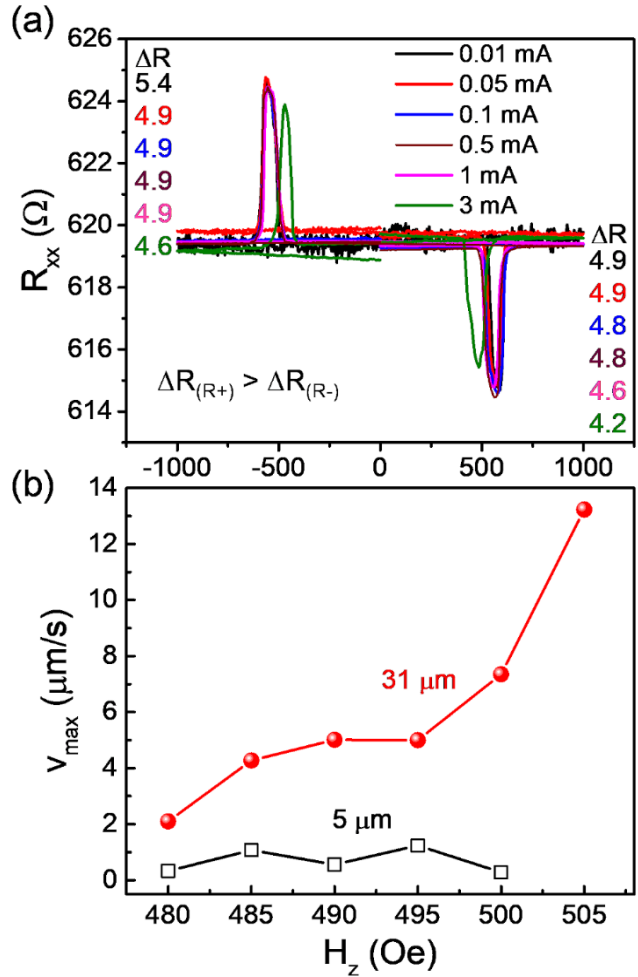

**FIG. 4.** (a) Asymmetric magnetoresistance for different sensing currents. For 3 mA sensing current, the peak appears at lower field is due to the Joule heating effect. The change in resistance ΔR is larger at $R_+$ peaks than the $R_-$. (b) Peak domain wall motion velocity ($v_{max}$) from the magnetization relaxation at various perpendicular fields. The relaxation curve was divided in two parts; the first part represents 5 μm distance, whereas, the second part represents the 31 μm distance.